\documentclass[]{PoS}
\usepackage{cite}
\usepackage{amsmath}
\newcommand{\ep}{\varepsilon}
\newcommand{\MS}{\overline{\sf MS}}
\newcommand{\Mvec}{\mbox{\rm\bf M}}
\newcommand{\Li}{\mbox{\rm Li}}

\newcommand{\gsim}{\raisebox{-0.07cm   }
{$\, \stackrel{>}{{\scriptstyle\sim}}\, $}}

\allowdisplaybreaks[4]

\title{
{\footnotesize 
DESY 14--121, DO-TH 14/12, MITP/14-047, SFB/CPP-14-36, LPN14-085}\\
3-Loop Heavy Flavor Corrections in Deep-Inelastic Scattering with Two Heavy Quark Lines}
\ShortTitle{DIS Contributions with Two Heavy Quark Lines}

\author{J.~Ablinger$^a$, J.~Bl{\"u}mlein$^b$, A.~De~Freitas$^b$, A.~Hasselhuhn$^{a,b}$, A.~von~Manteuffel$^c$, 
M.~Round$^{a,b}$, C.~Schneider$^a$ 
and \speaker{F.~Wi\ss{}brock}$^{a,b,d}$\thanks{
We would like to thank M.~Steinhauser for the possibility to use the code {\tt Q2e/EXP}. This work was supported 
in part by DFG Sonderforschungsbereich Transregio 9, Computergest\"utzte Theoretische Teilchenphysik, Studienstiftung 
des Deutschen Volkes, the Austrian Science Fund (FWF) grants P20347-N18 and SFB F50 (F5009-N15), the European 
Commission through contract PITN-GA-2010-264564 ({LHCPhenoNet}) and PITN-GA-2012-316704 ({HIGGSTOOLS}), by the 
Research Center ``Elementary Forces and Mathematical Foundations (EMG)'' of J. Gutenberg University Mainz and 
DFG, and by FP7 ERC Starting Grant  257638 PAGAP.}
\\
\llap{$^a$}~Research
Institute for Symbolic Computation (RISC), Johannes Kepler
University, Altenbergerstra\ss{}e 69, A-4040 Linz, Austria 
\\
\llap{$^b$}~Deutsches Elektronen--Synchrotron, DESY, Platanenallee 6, D--15738 Zeuthen, Germany
\\
\llap{$^c$}~PRISMA Cluster of Excellence, Institute of Physics, J.~Gutenberg University Mainz, D-55099 Mainz,Germany
\\
\llap{$^d$}~Institut des Hautes {\'E}tudes Scientifiques, IHES, Route de Chartres 35, \\ F--91440 Bures-sur-Yvette, France
\\
E-mail: \email{jablinge@risc.uni-linz.ac.at}, 
\email{Johannes.Bluemlein@desy.de},
\email{abilio.de.freitas@desy.de},
\email{alexander.hasselhuhn@desy.de},
\email{manteuffel@uni-mainz.de},
\email{mark.round@desy.de},
\email{cschneid@risc.uni-linz.ac.at},
\email{fabian.wissbrock@desy.de}
}

\abstract{
We consider gluonic contributions to the heavy flavor Wilson coefficients at 3-loop order in QCD with
two heavy quark lines in the asymptotic region $Q^2 \gg m_{1(2)}^2$. Here we report on the complete result
in the case of two equal masses $m_1 = m_2$ for the massive operator matrix element $A_{gg,Q}^{(3)}$, which 
contributes to the corresponding heavy flavor transition matrix element in the variable flavor number
scheme. Nested finite binomial sums and iterated integrals over square-root valued alphabets emerge in the
result for this quantity in $N$ and $x$-space, respectively. We also present results for the case of two 
unequal masses for the flavor non-singlet OMEs and on the scalar integrals ic case of $A_{gg,Q}^{(3)}$, which 
were calculated without a further approximation. The graphs can be expressed by finite nested binomial sums 
over generalized harmonic sums, the alphabet of which contains rational letters in the ratio $\eta = m_1^2/m_2^2$.}

\FullConference{Loops and Legs in Quantum Field Theory - LL 2014,\\
April 27th- March 2nd 2014\\
Weimar, Germany}

\begin{document}
\section{Introduction}
\label{sec:1}

\vspace*{1mm}
\noindent
Beginning with 3-loop order in QCD the heavy flavor contributions to the deep-inelastic structure functions, such as 
$F_2(x,Q^2)$, contain Feynman diagrams with two massive quark lines, either of equal or unequal mass. In the asymptotic
region of large virtualities $Q^2 \gg m^2$ for the former case, calculations of a series of Mellin moments \cite{Bierenbaum:2009mv} 
have been performed. It is well-known that the asymptotic picture agrees with the exact one at an accuracy of $O(1\%)$ at next-to-leading 
order for scales $Q^2/m^2 \gsim 10$ for the structure function $F_2(x,Q^2)$ \cite{Buza:1995ie}. By now, the complete set 
of logarithmic contributions \cite{Behring:2014eya} is known and the massive Wilson coefficients and/or operator matrix elements 
(OMEs) $L_{q,2}^{(3),\rm PS}, L_{g,2}^{(3)}$~\cite{Ablinger:2010ty,Behring:2014eya}, 
$L_{q,2}^{(3),\rm NS}$~\cite{Ablinger:2014vwa}, $H_{q,2}^{(3),\rm PS}$~\cite{Ablinger:14a}, 
$A_{qg,Q}^{(3)}, A_{qq,Q}^{(3), \rm PS}$~\cite{Ablinger:2010ty}, $A_{gq,Q}^{(3)}$~\cite{Ablinger:2014lka},
$A_{qq,Q}^{(3), \rm NS}$~\cite{Ablinger:2014vwa}, and $A_{Qq}^{(3), \rm PS}$~\cite{Ablinger:14a} have been calculated
completely.\footnote{For the notation we refer to Ref.~\cite{Bierenbaum:2009mv}.}

Since the ratio $m_c^2/m_b^2 \sim 1/10$, with $m_c$ and $m_b$ the charm and bottom quark masses, is not small, the charm quark 
cannot be considered massless at the scale $\mu = m_b$. Therefore a sequential decoupling of these two heavy quark flavors 
using the single quark decoupling in the usual variable heavy flavor scheme 
(VFNS)~\cite{Buza:1996wv,Bierenbaum:2009zt,Bierenbaum:2009mv} is not possible\footnote{The value of the decoupling scale in 
the VFNS is process dependent and usually {\it not} the scale of the decoupling heavy quark mass, cf.~Ref.~\cite{Blumlein:1998sh}.}.
Instead a generalization of the VFNS, decoupling two massive quarks at the same time, is necessary \cite{JB14a}.
  
In the case of two different masses $m_c$ and $m_b$, one may derive the asymptotic heavy 
flavor Wilson coefficients as well by considering the region of virtualities 
$Q^2 \gg m_b^2, m_c^2$, where power corrections $O(m_{c,b}^2/Q^2)$ can be neglected. The heavy flavor Wilson coefficients 
are known to factorize into the light flavor coefficients $C_{i,(2,L)}$ and the process--independent massive operator matrix 
elements (OMEs) $A_{ij}$, see  Refs.~\cite{Buza:1995ie,Buza:1996wv} for the corresponding relations.

The massive OMEs are evaluated in terms of Feynman diagrams with additional Feynman rules for the composite operator insertions. 
The moments $N = 2, 4, 6$ to all operator matrix elements have been calculated up to $O\left(\frac{m_c^6}{m_b^6} 
\ln^3\left(m_c^2/m_b^2\right)\right)$ for the contributions containing both charm and bottom-lines  
\cite{Ablinger:2011pb,Ablinger:2012qj,JB14a} using the projection method \cite{Bierenbaum:2009mv} through which the OMEs are mapped 
to tadpoles. They have been calculated using the code {\tt Q2e/EXP} \cite{Q2EXP}. This required a total computation time of about 1 CPU-year.

In this note we report on recent results on contributions to OMEs containing two massive fermion lines of equal and unequal 
quark masses. In Section~\ref{sec:2} we discuss the contributions of $O(\alpha_s^3 T_F^2 C_{F(A)})$ to the OME
$A_{gg,Q}$ in the case of equal masses. The renormalization of the unequal mass case is briefly discussed in 
Section~\ref{sec:3}. There we also present results for the massive OME in the flavor non-singlet and transversity cases
and for a scalar diagram with unequal masses at general Mellin variable $N$
and arbitrary mass ratio contributing to $A_{gg,Q}$. Section~\ref{sec:4} contains the conclusions. 
\section{The \boldmath{$O(\alpha_s^3 T_F^2)$} Contributions to $A_{gg,Q}$}
\label{sec:2}

\vspace*{1mm}
\noindent \normalsize
The contributions of $O(\alpha_s^3 T_F^2 C_{F,A})$ to the operator matrix element $A_{gg,Q}$ have been calculated in
Ref.~\cite{Ablinger:2014uka}. Here the color factors are $C_F = (N_c^2-1)/(2 N_c), C_A = N_c, T_F = 1/2$ in $SU(N_c)$ and
$N_c = 3$ for QCD. Most of the diagrams have been computed directly using Mellin-Barnes representations
and generating function techniques directly, i.e. including the corresponding numerator structures. This led to
large amounts of nested sums which were treated with the summation technologies encoded in the package {\tt Sigma} 
\cite{SIG1,SIG2}, based on advanced symbolic summation algorithms in the setting of difference fields 
\cite{Karr:81,Schneider:01,Schneider:05a,Schneider:07d,Schneider:08c,Schneider:10a,Schneider:10b,Schneider:10c,
Schneider:13b}, and the packages {\tt EvaluateMultiSums}, {\tt SumProduction} \cite{EMSSP}, and {\tt RhoSum}
\cite{ROUND}, which are all based on {\tt Sigma}. In part of the sums we used integration-by-parts reduction applying 
{\tt Reduze2} 
\cite{Studerus:2009ye,vonManteuffel:2012np}\footnote{The package {\tt Reduze2} uses {\tt Fermat} \cite{FERMAT} and {\tt
GiNac} \cite{Bauer:2000cp}.} 
and calculated the corresponding master integrals using differential equations and 
also applying Mellin-Barnes techniques.

As a main result of Ref.~\cite{Ablinger:2014uka} we present 
the constant part to the unrenormalized operator matrix element $a_{gg,Q;\rm T_F^2}^{(3)}$. It is finally obtained
by the following compact expression
\begin{eqnarray}
\label{eq:agg}
\lefteqn{a_{gg,Q;\rm T_F^2}^{(3)}(N) =} \nonumber\\ && {C_F T_F^2}
\Biggl\{
 \frac{16}{27} F S_1^3
+\frac{16 Q_4}{27 (N-1) N^3 (N+1)^3 (N+2)} S_1^2
+\Biggl[-\frac{16}{3} F S_2
\nonumber\\ &&
-\frac{32 Q_{10}}{81 (N-1) N^4 (N+1)^4 (N+2) (2 N-3) (2 N-1)} \Biggr] S_1
-\frac{16 Q_4}{9 (N-1) N^3 (N+1)^3 (N+2)} S_2
\nonumber\\ &&
-\frac{2 Q_{13}}{243 (N-1) N^5 (N+1)^5 (N+2) (2 N-3) (2 N-1)}
- F \left[ \frac{352}{27}   S_3
- \frac{64}{3}  S_{2,1} \right]
\nonumber\\ &&
+\Biggl[\frac{16}{3} F S_1
-\frac{8 Q_8}{9 (N-1) N^3 (N+1)^3 (N+2)}\Biggr] \zeta_2
+\frac{Q_3}{9 (N-1) N^2 (N+1)^2 (N+2)} \zeta_3
\nonumber\\ &&
-\binom{2N}{N} 
\frac{16 Q_5}{3(N-1) N (N+1)^2 (N+2) (2 N-3) (2 N-1)} \frac{1}{4^N}\left(\sum_{i=1}^N \frac{4^i S_1(i-1)}{i^2 \binom{2i}{i}} - 
7 
\zeta_3\right) 
\Biggr\}
\nonumber\\ &&
+{C_A T_F^2} 
\Biggl\{
-\frac{4 Q_2}{135 (N-1) N^2 (N+1)^2 (N+2)} S_1^2
+\frac{16 \big(4 N^3+4 N^2-7 N+1\big)}{15 (N-1) N (N+1)} [S_{2,1} - S_3]
\nonumber\\ &&
+\frac{Q_{12}}{3645(N-1) N^4 (N+1)^4 (N+2) (2 N-3) (2 N-1)} 
\nonumber\\ &&
-\frac{8 Q_{11}}{3645 (N-1) N^3 (N+1)^3 (N+2) (2 N-3) (2 N-1)} S_1
+\frac{4 Q_7}{135 (N-1) N^2 (N+1)^2 (N+2)} S_2
\nonumber\\ &&
-\binom{2N}{N} 
\frac{4 Q_9}{45 (N-1) N (N+1)^2 (N+2) (2 N-3) (2 N-1)} \frac{1}{4^N} \left(\sum_{i=1}^N 
\frac{4^i S_1(i-1)}{i^2 \binom{2i}{i}} -7 \zeta_3\right)
\nonumber\\ &&
+\Biggl[\frac{4 Q_6}{27 (N-1) N^2 (N+1)^2 (N+2)}
-\frac{560}{27} S_1
\Biggr] \zeta_2
+\Biggl[-\frac{7 Q_1}{270 (N-1) N (N+1) (N+2)}
\nonumber\\ &&
-\frac{1120}{27} S_1 \Biggr] \zeta_3 \Biggr\}~.
\end{eqnarray}
Here we define
\begin{eqnarray}
F(N) = \frac{(2+N+N^2)^2}{(N-1) N^2 (N+1)^2 (N+2)} \equiv F,
\end{eqnarray}
and $Q_i$ denote polynomials in $N$, cf.~\cite{Ablinger:2014uka}.
The equal mass contribution to the massive OME $A_{gg,Q}^{(3)}$ can be described by harmonic sums 
\cite{Vermaseren:1998uu,Blumlein:1998if}  $S_{\vec{a}}(N) \equiv S_{\vec{a}}$,
\begin{eqnarray}
S_{b,\vec{a}}(N) = \sum_{k = 1}^N \frac{({\rm sign}(b))^k}{k^{|b|}} S_{\vec{a}}(k),~~~S_\emptyset = 1, b, a_i \in \mathbb{Z} 
\backslash \{0\}
\end{eqnarray}
and one inverse binomial sum weighted by a harmonic sum \cite{Ablinger:2014bra}. In $x$-space the latter sum 
results in an iterated integral of square root-valued letters, extending the space of functions for contributions to 
the 3-loop Wilson coefficients having been know so far for the first time. 
\section{Scalar Integrals of the \boldmath{$O(\alpha_s^3 T_F^2)$} Contributions to \boldmath{$A_{gg,Q}$} with Two Different 
Masses}
\label{sec:3}

\vspace*{1mm}
\noindent
We have calculated all scalar integrals contributing to the OME $A_{gg,Q}$ in case of two different masses both in
$x$- and $N$-space, without any approximation, like e.g. an expansion in $\eta = m_c^2/m_b^2$. We first describe the main steps
of the renormalization of these contributions, cf. Ref.~\cite{JB14a} for details, present results in the flavor non-singlet 
case, and finally give an example for the scalar integrals contributing to $A_{gg,Q}^{(3)}$. 
\subsection{Renormalization}
\label{sec:3.1}

\vspace*{1mm}
\noindent
The renormalization of the operator matrix elements with two heavy quark flavors is performed as a generalization of the 
single mass case in Ref.~\cite{Bierenbaum:2009mv}. Here the sub-set of graphs with two massive fermion lines is considered.
It is technically of advantage to treat the equal and different mass cases together. The renormalization of the unequal mass 
case is then obtained by subtracting the contributions in the equal mass case taken from \cite{Bierenbaum:2009mv}. The quark mass
is either renormalized in the on-shell renormalization scheme or the $\overline{\sf MS}$ scheme~\cite{MASS}. The factorization 
relation \cite{Buza:1995ie} at large virtualities strictly requires 
the external legs of the operator matrix elements to be on--shell. Charge renormalization is easiest carried out in a {\sf MOM}--scheme 
applying the background field method \cite{BGF}. Afterwards a finite renormalization to the $\MS$--scheme is performed.
The remaining ultra-violet singularities of the composite operators are renormalized 
via the corresponding $Z_{ij}$--factors and in a final step the collinear singularities are removed as they are 
absorbed into the parton distribution functions. In all the quantities, corrections due to the two mass scales $m_c$ and 
$m_b$ are emerging. Accordingly one may generalize the VFNS w.r.t. the simultaneous decoupling of both $m_c$ and $m_b$,
which is needed from 3-loop order onwards, since here graphs with two heavy quark lines contribute. Their mass dependence does 
not factorize, unlike the case up to 2-loop order. 
\subsection{Operator matrix elements with $m_1 \neq m_2$}
\label{sec:3.2}

\vspace*{1mm}
\noindent
In the case of a general Mellin variable $N$ the 3-loop graphs with two different masses have been 
calculated for the OME in the non-singlet and transversity cases and for general mass ratios $\eta$.
Only a few Feynman diagrams contribute in these cases. Here we show the unequal mass contribution to the constant 
part of the unrenormalized OME, cf. \cite{Bierenbaum:2009mv}, for the non-singlet and transversity cases
\begin{eqnarray}
a_{qq,Q}^{(3), \rm{NS}}&=&
 C_F T_F^2 \Biggl\{
 \left(\frac{32}{27} S_1-\frac{8 \left(3 N^2+3 N+2\right)}{27 N (N+1)}\right) \ln^3(\eta)
 +\Biggl[-\frac{R_1}{18 N^2 (N+1)^2 \eta}
 \nonumber\\&&
 +\Bigl[
 \frac{\left(3 N^2+3 N+2\right) (\eta+1) \left(5 \eta^2+22 \eta+5\right)}{36 N (N+1) \eta^{3/2}}
 -\frac{(\eta+1) \left(5 \eta^2+22 \eta+5\right)}{9 \eta^{3/2}} S_1\Bigr] \ln \left(\frac{1+\eta_1}{1-\eta_1}\right)
 \nonumber\\&&
 +\frac{2 \left(5 \eta^2+2 \eta+5\right)}{9 \eta} S_1
 +\ln(1-\eta) \left(\frac{16 \left(3 N^2+3 N+2\right)}{9 N (N+1)}-\frac{64}{9} S_1\right)+\frac{32}{9} S_2\Biggr] \ln^2(\eta)
 \nonumber\\&&
 +\Biggl[\frac{40 (\eta-1)(\eta+1)}{9 \eta} S_1
 -\frac{10 \left(3 N^2+3 N+2\right) (\eta-1) (\eta+1)}{9 N (N+1) \eta}
 + \frac{(\eta+1) \left(5 \eta^2+22 \eta+5\right)}{9 \eta^{3/2}} 
 \nonumber\\&&
\times \Bigl[
 8 S_1
 -\frac{2 \left(3 N^2+3 N+2\right)}{N (N+1)}\Bigr] \Li_2\left(\eta_1\right)
 \nonumber\\&&
 +
 \frac{\left(\eta_1+1\right)^2  \left(-10 \eta^{3/2}+5 \eta^2+42 \eta-10 \eta_1+5\right)}{9 \eta^{3/2}}
 \left[
 \frac{\left(3 N^2+3 N+2\right) }{2 N (N+1)}
 -2  S_1
 \right] \Li_2(\eta)
 \Biggr] \ln(\eta)
 \nonumber\\&&
 +\frac{16 \left(3 N^4+6 N^3+47 N^2+20 N-12\right) \zeta_2}{27 N^2 (N+1)^2}
 +\frac{(\eta+1) \left(5 \eta^2+22 \eta+5\right)}{9 \eta^{3/2}}
 \Bigl[
 \frac{4 \left(3 N^2+3 N+2\right) }{N (N+1)}
 \nonumber\\&&
 -16 S_1\Bigr] \Li_3\left(\eta_1\right)
 +
 \frac{\left(\eta_1+1\right)^2 \left(-10 \eta^{3/2}+5 \eta^2+42 \eta-10 \eta_1+5\right)}{9 \eta^{3/2}}
 \Bigl[
 {2  S_1}
 \nonumber\\&&
 -\frac{\left(3 N^2+3 N+2\right) }{2 N (N+1)}\Bigr] \Li_3(\eta)
 +\left[\frac{16 \left(405 \eta^2-3238 \eta+405\right)}{729 \eta}+\frac{256 \zeta_3}{27}-\frac{640 \zeta_2}{27}\right] S_1
 \nonumber\\&&
 +\left[\frac{128 \zeta_2}{9}+\frac{3712}{81}\right] S_2
 -\frac{1280}{81} S_3+\frac{256}{27} S_4-\frac{64 \left(3 N^2+3 N+2\right) \zeta_3}{27 N (N+1)}
 -\frac{4 R_2}{729 N^4 (N+1)^4 \eta}
 \Biggr\}
\nonumber\\
\end{eqnarray}
and
\begin{eqnarray}
a_{qq,Q}^{(3),\rm{NS},\rm TR}&=&
C_F T_F^2 \Biggl\{
\left[\frac{16}{27} S_1-\frac{4}{9}\right] \ln^3(\eta)
+\Biggl[
-\frac{(\eta+5) (5 \eta+1)}{12 \eta}
+
\frac{(\eta+1) \left(5 \eta^2+22 \eta+5\right)} {72 \eta^{3/2}}
\left[4 S_1
-3\right] 
\nonumber\\&&
\times
\ln\left(\frac{1-\eta_1}{1+\eta_1}\right)
+\frac{\left(5 \eta^2+2 \eta+5\right)}{9 \eta} S_1
+\ln(1-\eta) \left(\frac{8}{3}-\frac{32}{9} S_1\right)
+\frac{16}{9} S_2
\Biggr] \ln^2(\eta)
\nonumber\\&&
+\Biggl[
\frac{20 (\eta-1)(\eta+1)}{9 \eta} S_1
-\frac{5 (\eta-1) (\eta+1)}{3 \eta}
+
\frac{(\eta+1) \left(5 \eta^2+22 \eta+5\right)}{9 \eta^{3/2}}
\left[4  S_1
-3\right] 
\nonumber\\&&
\times \Li_2\left(\eta_1\right)
+\frac{\left(\eta_1+1\right)^2 \left(-10 \eta^{3/2}+5 \eta^2+42 \eta-10 \eta_1+5\right)} {36 \eta^{3/2}}
\left[
3 
-4 S_1\right] \Li_2(\eta)\Biggr] \log (\eta)
\nonumber\\&&
+\frac{8 \zeta_2}{9}
+\frac{(\eta+1) \left(5 \eta^2+22 \eta+5\right)}{9 \eta^{3/2}} \left[
{6 }
-{8  S_1}\right] \Li_3\left(\eta_1\right)
\nonumber\\&&
+
\frac{\left(\eta_1+1\right)^2 \left(-10 \eta^{3/2}+5 \eta^2+42 \eta-10 \eta_1+5\right)}{36 \eta^{3/2}}
\left[
 4 S_1
-3 \right] \Li_3(\eta)
\nonumber\\&&
+\left(\frac{8 \left(405 \eta^2-3238 \eta+405\right)}{729 \eta}+\frac{128 \zeta_3}{27}-\frac{320 \zeta_2}{27}\right) S_1
+\left(
\frac{64 \zeta_2}{9}+\frac{1856}{81}\right) S_2
\nonumber\\&&
-\frac{640}{81} S_3
+\frac{128}{27} S_4-\frac{32 \zeta_3}{9}
-\frac{2 R_3}{243 N^2 (N+1)^2 \eta}\Biggr\}~.
\end{eqnarray}
Here $\eta_1 = \sqrt{\eta}$ and the polynomials $R_i$ read
\begin{eqnarray}
 R_1&=&15 \eta^2 N^4+78 \eta N^4+15 N^4+30 \eta^2 N^3+156 \eta N^3+30 N^3+25 \eta^2 N^2+18 \eta N^2+25 N^2
 \nonumber\\&&+10 \eta^2 N+4 \eta N+10 N+32 \eta
 \\
 R_2&=&1215 \eta^2 N^8-1596 \eta N^8+1215 N^8+4860 \eta^2 N^7-6384 \eta N^7+4860 N^7+8100 \eta^2 N^6
 \nonumber\\&&
 -25844 \eta N^6+8100 N^6
 +7290 \eta^2 N^5-39348 \eta N^5+7290 N^5+3645 \eta^2 N^4-20304 \eta N^4
 \nonumber\\&&
 +3645 N^4+810 \eta^2 N^3-140 \eta N^3+810 N^3+432 \eta N^2+288 \eta N+864 \eta
\\
 R_3&=&405 \eta^2 N^4-532 \eta N^4+405 N^4+810 \eta^2 N^3-1064 \eta N^3+810 N^3+405 \eta^2 N^2-1012 \eta N^2
 \nonumber\\&&
 +405 N^2+96 \eta N+288 \eta~.
\end{eqnarray}
The $N$ dependence is described by rational contributions and harmonic sums, while for the dependence on the mass-ratio 
also polylogarithms \cite{LEWIN} arise.
\begin{figure}[!htb]
\begin{center}
\includegraphics[scale=1.0]{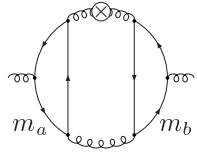}
\end{center}
\caption{A typical scalar diagram contributing to the OME $A_{gg,Q}^{(3)}$.
It is symmetric under the change of the  mass assignment $m_a \leftrightarrow m_b$. The diagram has been
drawn using {\tt Axodraw}~\cite{Vermaseren:1994je}. 
\label{D7}}
\end{figure}

Next we consider the contributions due to unequal masses for the gluonic  operator matrix element $A_{gg,Q}^{(3)}$. 
We have obtained the $x$- and Mellin space representations for all contributing scalar diagrams. The calculation of the 
Feynman diagrams is performed as follows. The Feynman parameter integrals can be carried out by introducing an intermediary 
Mellin-Barnes representation \cite{MELB}. The corresponding sum of residues does usually not converge in the complete 
integration region 
for the Feynman parameter integrals. In order to ensure convergence, several transformations of the integrals have to be 
performed. The integration domain of the final integral is then split into two regions for each of which a convergent sum of 
residues can be obtained. The sums are solved applying the algorithms encoded 
in the package {\tt Sigma} \cite{SIG1,SIG2}, {\tt EvaluateMultiSums} and 
{\tt SumProduction} \cite{EMSSP}.
We map the final integral to obtain the form of a Mellin transform 
\begin{eqnarray}
\Mvec[f(x)](N) = \int_0^1 dx~x^N~f(x)~
\end{eqnarray}
and rewrite the $x$--space representation $f(x)$ in terms of iterated integrals with argument $x$ by using techniques 
inspired by \cite{Brown:2008um}. The result cannot be described within the class of hyperlogarithms only, but  more 
general iterated integrals involving square-root valued integration kernels~\cite{Ablinger:2014bra} occur, see also 
\cite{BHAB}.
In the corresponding Mellin space expressions, finite nested binomial sums weighted by $\binom{2k}{k}^{\pm 1}$ are 
emerging. Here also the function 
\begin{eqnarray}
 L_1(\eta)&=&\frac{1}{2} \int_0^\eta dx \frac{\sqrt{x}}{1-x} \ln^2(x) \nonumber\\ &=&
 4 \Biggl\{\Li_3(\eta_1) - \Li_3(-\eta_1) + \ln(\eta_1) 
\left[\Li_2(-\eta_1) - \Li_2(\eta_1)\right] - \frac{1}{2} \ln^2(\eta_1) 
\ln\left(\frac{1-\eta_1}{1+\eta_1}\right)
\nonumber\\ &&
- 2 \eta_1 \left[1  - \ln(\eta_1) +  \frac{1}{2} \ln^2(\eta_1)\right]\Biggr\},
\end{eqnarray}
is frequently observed.
 
In a final step we apply the computer algebra package {\tt HarmonicSums} 
\cite{Ablinger:2010kw,Ablinger:2011te,Ablinger:2013cf,Ablinger:2013hcp} to generate a difference equation for 
the Mellin transform of $f(x)$ and solve it by using the packages {\tt Sigma}, {\tt EvaluateMultiSums} and  {\tt 
SumProduction}. As an example we present the $N$--space result for the diagram in Figure~\ref{D7}
\normalsize

{\footnotesize
 \begin{eqnarray}
\lefteqn{I(N) =} \\ &&
 \left(m_b^2\right)^{-3+3/2 \ep} 
\left[\frac{1+(-1)^N}{2}\right]
\Biggl\{
-\frac{\eta +1}{24 \ep \eta^2 (N+1)}
+
\Biggl[
P_8 \frac{1}{5760 \eta ^3 N (N+1)^2 (N+2)}
\nonumber\\&&
+\frac{1}{45} \frac{2^{-2 N-9} \binom{2 N}{N} P_2}{(\eta -1) \eta ^3 (N+1)^2 (N+2)}
\sum_{i_1=1}^N \frac{2^{2 i_1} 
\big(\frac{\eta}{-1+\eta }\big)^{i_1}}
{\binom{2 i_1}{i_1}} 
-\frac{1}{45} \frac{2^{-2 N-8} \binom{2 N}{N}}{\eta ^2 (N+1)^2 (N+2)} P_4
\nonumber\\&&
+\frac{\big(\frac{\eta }{\eta -1}\big)^N}{11520 (\eta -1) \eta ^2 N (N+1)^2 (N+2)} P_5
+\frac{(1-\eta )^{-N} P_6}{11520 (\eta -1) \eta^3 N (N+1)^2 (N+2)}
+\frac{S_1\left(\frac{1}{1-\eta},N\right)}{360 (N+1)}
+\frac{S_1\left(\frac{\eta }{\eta -1},N\right)}{360 \eta ^3 (N+1)} 
\nonumber\\
&&
-\frac{1}{45} P_{11} \frac{2^{-2 N-9} \binom{2 N}{N}}{(\eta -1) \eta (N+1)^2 (N+2)} 
\sum_{i_1=1}^N \frac{2^{2 i_1} (1-\eta)^{-i_1}}
{\binom{2 i_1}{i_1}}\Biggr] \log^2(\eta)
+\Biggl[
-\frac{P_7}{5760 \eta ^3 (N+1)^2 (N+2)}
\nonumber\\ &&
+\frac{1}{45} (\eta +1) \frac{2^{-2 N-7} \binom{2 N}{N}}{\eta ^3 (N+1)^2 (N+2)} P_2
-\frac{1}{45} \frac{2^{-2 N-8} \binom{2 N}{N} P_2}{(\eta -1) \eta ^3 (N+1)^2 (N+2)}
\sum_{i_1=1}^N \frac{2^{2 i_1}
\big(\frac{\eta}{-1+\eta }\big)^{i_1} S_1\left(\frac{-1+\eta }{\eta},i_1\right)}{\binom{2 i_1}{i_1}} 
\nonumber\\&&
+\frac{1}{90} \big(\eta ^3-1\big) \frac{1}{\eta ^3 N (N+1)^2 (N+2)} S_1\left(N\right)
+\frac{(1-\eta )^{-N}}{5760 (\eta -1) \eta ^3 N (N+1)^2 (N+2)} P_6 S_1\left(1-\eta ,N\right)
\nonumber\\&&
-\frac{\big(\eta ^3-1\big) S_2\left(N\right)}{180 \eta ^3 (N+1)}
-\frac{\big(\frac{\eta }{\eta-1}\big)^N}{5760 (\eta -1) \eta ^2 N (N+1)^2 (N+2)} 
P_5 S_1\left(\frac{\eta-1}{\eta },N\right)
+\frac{S_{1,1}\left(\frac{1}{1-\eta },1-\eta,N\right)}{180 (N+1)}
\nonumber\\&&
-\frac{1}{180} \frac{1}{\eta ^3 (N+1)} S_{1,1}\left(\frac{\eta }{\eta -1},\frac{\eta -1}{\eta },N\right)
\nonumber\\&&
-\frac{1}{45} P_{10} \frac{2^{-2 N-8} \binom{2 N}{N}}
{(\eta -1) \eta  (N+1)^2 (N+2)} \sum_{i_1=1}^N \frac{2^{2 i_1} (1-\eta )^{-i_1} S_1\left(1-\eta ,i_1\right)}
{\binom{2 i_1}{i_1}}\Biggr] 
\log (\eta )
\nonumber\\ &&
+ 
\Bigl[
\frac{\big(27 \eta ^2+10 \eta+27\big)}{5760 \eta ^{5/2} (N+1)} 
-\frac{2^{-2 N-8} \binom{2 N}{N} P_1}{45 \eta ^{5/2} (N+1)^2 (N+2)} \Bigr]
L_1(\eta)
-\frac{1}{45} (\eta -1) \frac{2^{-2 N-6} \binom{2 N}{N}}{\eta ^3 (N+1)^2 (N+2)} P_2
\nonumber\\&&
+\frac{2^{-2 N} \binom{2 N}{N} P_2}{11520 (\eta -1) \eta ^3 (N+1)^2 (N+2)}
\sum_{i_1=1}^N \frac{2^{2 i_1} \big(\frac{\eta }{-1+\eta }\big)^{i_1} 
\Bigl[S_{1,1}\left(\frac{-1+\eta }{\eta },1,i_1\right)
-S_2\left(\frac{-1+\eta }{\eta },i_1\right)\Bigr]
}{\binom{2 i_1}{i_1}}
\nonumber\\&&
-\frac{1}{2880 \eta ^3 N (N+1)^2 (N+2)} P_9
+\frac{(\eta +1) P_3 S_1\left(N\right)}{5760 \eta ^3 (N+1)^2 (N+2)}
+ \frac{\big(\eta ^3+1\big)}{180 \eta ^3 N (N+1)^2 (N+2)} \Bigl[S_2\left(N\right)-S_1^2(N)\Bigr]
\nonumber\\&&
+\frac{\big(\eta^3+1\big) S_3(N)}{180 \eta ^3 (N+1)}
+\frac{(1-\eta )^{-N} P_6}{5760 (\eta -1) \eta ^3 N (N+1)^2 (N+2)}
\Bigl[S_{1,1}\left(1-\eta ,1,N\right)-S_2\left(1-\eta ,N\right)\Bigr]
\nonumber\\&&
+\frac{\big(\frac{\eta }{\eta -1}\big)^N P_5}{5760 (\eta -1) \eta ^2 N (N+1)^2 (N+2)} 
\Bigl[S_{1,1}\left(\frac{\eta -1}{\eta},1,N\right)
-S_2\left(\frac{\eta -1}{\eta},N\right)\Bigr]
+\frac{1}{180(N+1)}
\Bigl[
S_1\left(\frac{1}{1-\eta},N\right) 
\nonumber\\&&
S_{1,1}\left(1-\eta ,1,N\right)
-S_{1,2}\left(\frac{1}{1-\eta},1-\eta ,N\right)
+S_{1,2}\left(1-\eta ,\frac{1}{1-\eta},N\right)
-S_{1,1,1}\left(1-\eta ,1,\frac{1}{1-\eta},N\right)
\nonumber\\&&
-S_{1,1,1}\left(1-\eta ,\frac{1}{1-\eta},1,N\right)
\Bigr]
+\frac{1}{180 \eta ^3 (N+1)} \Bigl[
S_1\left(\frac{\eta }{\eta -1},N\right) S_{1,1}\left(\frac{\eta -1}{\eta},1,N\right)
\nonumber\\&&
+S_{1,2}\left(\frac{\eta -1}{\eta },\frac{\eta}{\eta -1},N\right)
-S_{1,2}\left(\frac{\eta }{\eta-1},\frac{\eta -1}{\eta },N\right)
-S_{1,1,1}\left(\frac{\eta -1}{\eta },1,\frac{\eta }{\eta-1},N\right)
\nonumber\\&&
-S_{1,1,1}\left(\frac{\eta -1}{\eta},\frac{\eta }{\eta -1},1,N\right)
\Bigr]
+\frac{2^{-2 N} \binom{2 N}{N} P_{10} }{11520 (\eta -1) \eta  (N+1)^2 (N+2)} 
\nonumber\\ && \times
\sum_{i_1=1}^N \frac{2^{2 i_1} (1-\eta)^{-i_1} \Bigl[S_2\left(1-\eta ,i_1\right)
-S_{1,1}\left(1-\eta,1,i_1\right)\Bigr]}{\binom{2 i_1}{i_1}}
\Biggr\}~,
\nonumber
\end{eqnarray}}

\noindent
\normalsize
where the $P_i$ represent different polynomials in the variables $N$ and $\eta$. Additionally to harmonic 
sums~\cite{Vermaseren:1998uu,Blumlein:1998if}, we also observe generalized harmonic sums\cite{Moch:2001zr,Ablinger:2013cf} 
\begin{eqnarray}
S_{b,\vec{a}}(c,\vec{d},N) = \sum_{k = 1}^N \frac{c^k}{k^{b}} S_{\vec{a}}(\vec{d},k),~~~S_\emptyset = 1,~~~b, a_i \in 
\mathbb{N} \backslash \{0\},~~~c, d_i \in \mathbb{R} \backslash \{0\}.
\end{eqnarray}
The real numerator weights are partly rational functions of the ratio $\eta$. Furthermore, nested binomial 
sums over generalized sums contribute \cite{Ablinger:2013eba,Ablinger:2014yaa,Ablinger:2014bra}. Usually, the expansion of the 
integrals like $I(N)$
in terms of $\eta \ll 1$ up to a finite order for general values of $N$ is not possible in general. However, if a fixed integer
value is chosen for $N$, the expansion may be performed. Here one obtains the same result as calculating the corresponding
moment using the code {\tt Q2e/EXP} \cite{Q2EXP}.
\section{Conclusions}
\label{sec:4}

\vspace*{1mm}
\noindent \normalsize
We calculated contributions of $O(T_F^2 C_{F(A)})$ 
to the gluonic massive operator matrix elements at 3-loop order in QCD from graphs with two massive quark lines, both 
for equal and unequal heavy quark internal lines\footnote{The case of one vanishing quark mass has been dealt with
in Refs.~\cite{Ablinger:2010ty,Behring:2013dga} before.}. In the calculation of these diagrams the evaluation of Mellin-Barnes 
integrals usually requires to close the contour either to the left or the right, depending on the value of one of the Feynman 
parameters.  Both in 
intermediary and the final result, nested finite binomial sums occur, weighted by harmonic sums in the equal mass case.
For unequal masses also generalized finite harmonic sums are present, the letters of which are rational expressions of the mass 
ratio of the two quarks. For general values of the Mellin variable $N$ the expansion in the mass ratio is not straightforward.
One is rather advised to deal with the case of general mass ratios. The present summation technologies allow to compute the complete 
result, however.


\end{document}